\documentclass{aa}
\usepackage{graphicx}
\usepackage{times}
\usepackage{natbib}
\usepackage{booktabs} 
\usepackage{morefloats}
\usepackage[hyphenbreaks]{breakurl}
\usepackage{hyperref}
\usepackage{amsmath}
\usepackage{amssymb}
\usepackage{cleveref}

\def\apgt{\ {\raise-.5ex\hbox{$\buildrel>\over\sim$}}\ }
\def\aplt{\ {\raise-.5ex\hbox{$\buildrel<\over\sim$}}\ }
\def\lteq{\ {\raise-.5ex\hbox{$\buildrel<\over-$}}\ }
\def\gteq{\ {\raise-.5ex\hbox{$\buildrel>\over-$}}\ }

\newcommand{\MSun}{\mbox{${M}_\odot$}}

\newcommand{\MEarth}{\mbox{${M}_\oplus$}}

\def\apgt{\ {\raise-.5ex\hbox{$\buildrel>\over\sim$}}\ }
\def\aplt{\ {\raise-.5ex\hbox{$\buildrel<\over\sim$}}\ }
\def\lteq{\ {\raise-.5ex\hbox{$\buildrel<\over-$}}\ }

\newcommand{\OHc}{Oort-Hills cloud}
\newcommand{\HOc}{Oort-Hills cloud}

\newcommand{\OOc}{\"Opik-Oort cloud}

\def\aap{\ {A\&A}\ }

\def\aj{\ {AJ}\ }

\def\apj{\ {ApJ}\ }
\def\apjl{\ {ApJL}\ }
\def\apjs{\ {ApJS}\ }
\def\apss{\ {Ap\&SS}\ }
\def\araa{\ {ARA\&A}\ }

\def\bain{\ {Bul. Astron. Inst. Neth.}\ }

\def\grl{\ {Geoph. R. L.}\ }
\def\icarus{\ {Icarus}\ }

\def\mnras{\ {MNRAS}\ }

\def\nat{\ {Nat}\ }

\def\na{\ {New Astron.}\ }

\def\pasj{\ {Publ. Astr. Soc. Japan}\ }

\def\psj{\ {The Planetary Science J.}\ }

\spacing{1}

\begin{document}

\title{Oort Cloud Ecology. III. The Sun left the parent star cluster shortly after the
giant planets formed}
\titlerunning{Oort Cloud Ecology III}

\author{
Simon Portegies Zwart$^1$
and
Shuo Huang$^{1,2}$
}

\institute{
Observatory, Leiden University, PO Box 9513, 2300
RA, Leiden, The Netherlands
\and
Department of Astronomy, Tsinghua University, 100084 Beijing, China
}

\date{}

\abstract
    { The Sun was born in a clustered environment with $\aplt 10\,000$
      other
      stars~\citep{2009ApJ...696L..13P,2010ARA&A..48...47A,2015AJ....150...26H,2023A&A...670A.105A}.
      Being an isolated star today, the Sun must have left the
      nest. We do not directly know when that happened, how violent
      the ejection was, or how far the Solar siblings have drifted
      apart \citep{2009ApJ...696L..13P}. }
    {The mass of the fragile outer \OOc\, (between $r_{\rm inner} \sim
      30\,000$\,au and $200\,000$\,au from the Sun
      \citep{1932PAAAS..67..169O,1950BAN....11...91O}) and the orbital
      distribution of planetesimals in the inner \HOc\, (between $\sim
      1000$\,au and $\sim 30\,000$\,au \citep{1981AJ.....86.1730H})
      are sensitive to the dynamical processes involving the Sun in
      the parent cluster. We aim at understanding the extend to which
      observing the Oort cloud constrains the Sun's birth
      environment.}
    {This is achieved by a combination of theoretical arguments and N-body simulations.}
    {We show that the current mass of the \OOc\, (between
      0.2\,\MEarth\, and
      $2.0$\,\MEarth~\citep{2005ApJ...635.1348F,2022arXiv220600010K})
      is best explained if the Sun left the nest within $\sim 20$\,Myr
      after the giant planets formed and migrated. }
    {As a consequence, the possible dynamical encounter with another
      star carving the Kuiper belt
      \citep{2024NatAs...8.1380P,2024ApJ...972L..21P}, the Sun's
      abduction of Sedna \citep{2015MNRAS.453.3157J}, and other
      perturbations induced by nearby stars then must have happened
      shortly after the giant planets in the Solar system formed, but
      before the Sun left the parent cluster.  Signatures of the time
      spend in the parent cluster must still be visible in the outer
      parts of the Solar system today. The strongest constraints will
      be the discovery of a population of relatively low-eccentricity
      ($e \aplt 0.9$) inner Oort-cloud (but $500 \aplt a \aplt
      10^4$\,au) objects. }

    
\maketitle
    
\section{Introduction}

The Oort cloud contains $\sim 6 \times 10^{11}$ objects larger than
$1$\,km \citep{2005ApJ...635.1348F,2010Sci...329..187L} that were
ejected from the circumstellar disk into wide and highly eccentric
orbits through gravitational assist by the four giant planets
\citep{1980Icar...42..406F,1987AJ.....94.1330D,2007MNRAS.381..779E}.
Planetesimals are generally not ejected through a single interaction
but have many pericenter passages. Upon each return to pericenter, the
comet receives a kick from a giant planet pumping its eccentricity to
$e \apgt 0.999$ in $\aplt 10$\,Myr \citep{2004come.book..153D}. Once
reaching such a high eccentricity, the planetesimals' survival becomes
a random walk with $\propto n^{-1/2}$ probability of being kicked to
$e \apgt 0.9999$ in $n$ orbits \citep{1979MNRAS.187..445Y}.  Once the
semi-major axis exceeds approximately $10^5$\,au
\citep{1984Natur.311...38S} the planetesimal travels beyond the Sun's
Hill radius in the Galactic potential, and is lost.

A planetesimal may prevent escape if at apocenter distance $\apgt
10^4$\,au, tidal forces of nearby stars circularize its orbit
\citep{1986Icar...65...13H,1999Icar..137...84W}, causing the pericenter
to lift beyond Neptune's influence at $\sim 35$\,au
\citep{1987AJ.....94.1330D,2021ApJ...920..148B,2024MNRAS.527.3054H}. Both
processes operate on a timescale of $\sim 100$\,Myr and compete, with
eccentricity pumping being somewhat more effective in ejecting
planetesimals than perihelion lifting is in preserving them. Once the
disk becomes depleted and an insufficient reservoir of planetesimals
is left over to replenish the Oort cloud, nearby stars, and the
Galactic tidal field drives the Oort cloud's erosion on a timescale of
3000\,Myr \citep{1985AJ.....90.1548H} to 13\,000\,Myr
\citep{2018MNRAS.473.5432H}. Eventually, the Solar system loses its
\OOc, but the inner parts survive. The \HOc\, is protected against
such erosion by the Sun's potential well.

The competition between eccentricity pumping and Galactic tidal
circularization satisfactorily explains the eccentricities and
inclinations of the \OOc, derived from long-period comets
\citep{1999Icar..137...84W,2020CeMDA.132...43F}. The Sun, then, was not
born with an Oort cloud but grew it over time and is currently losing
it \citep{2006Icar..184...59B}.

\section{Oort cloud formation-efficiency for an isolated Sun}

We quantify the mass evolution of the \OOc\, by performing direct
N-body simulations of the Sun, the four giant planets, and a
population of planetesimals in a thin debris disk of test particles
between 1\,au and 42\,au along the ecliptic,
see\,\cref{Sect:Methods}. Results are presented in terms of
$\mu_{\textrm{\"OO}} \equiv m_{\textrm{\"OO}}/m_{\rm \textrm{dm}}$,
which is the ratio of the mass in the \OOc\, ($m_{\textrm{\"OO}}$) and
the initial mass of the debris disk $m_{\textrm{dm}}$ without the
giant planet's cores of $m_{\textrm{pc}} \aplt 100$\,\MEarth\, (see
~\cref{Sect:Methods}). The disk makes an angle of $60.2^\circ$ to the
Galactic plane. For convenience, we start with the Sun's current
Galactic orbital parameters and adopt a smooth Galactic potential. All
these conditions were different upon the Sun's birth.  The planets,
for example, grew over a timescale of several Myr until they settled
into their current orbits
\citep{1984Icar...58..109F,brasser2024terrestrialplanetformationgiant}.
Although important for the general understanding of the Solar system,
these processes affect the formation of the Oort cloud only to second
order (see\,\cref{Sect:Methods}).

\begin{figure}
\center
\includegraphics[width=\columnwidth]{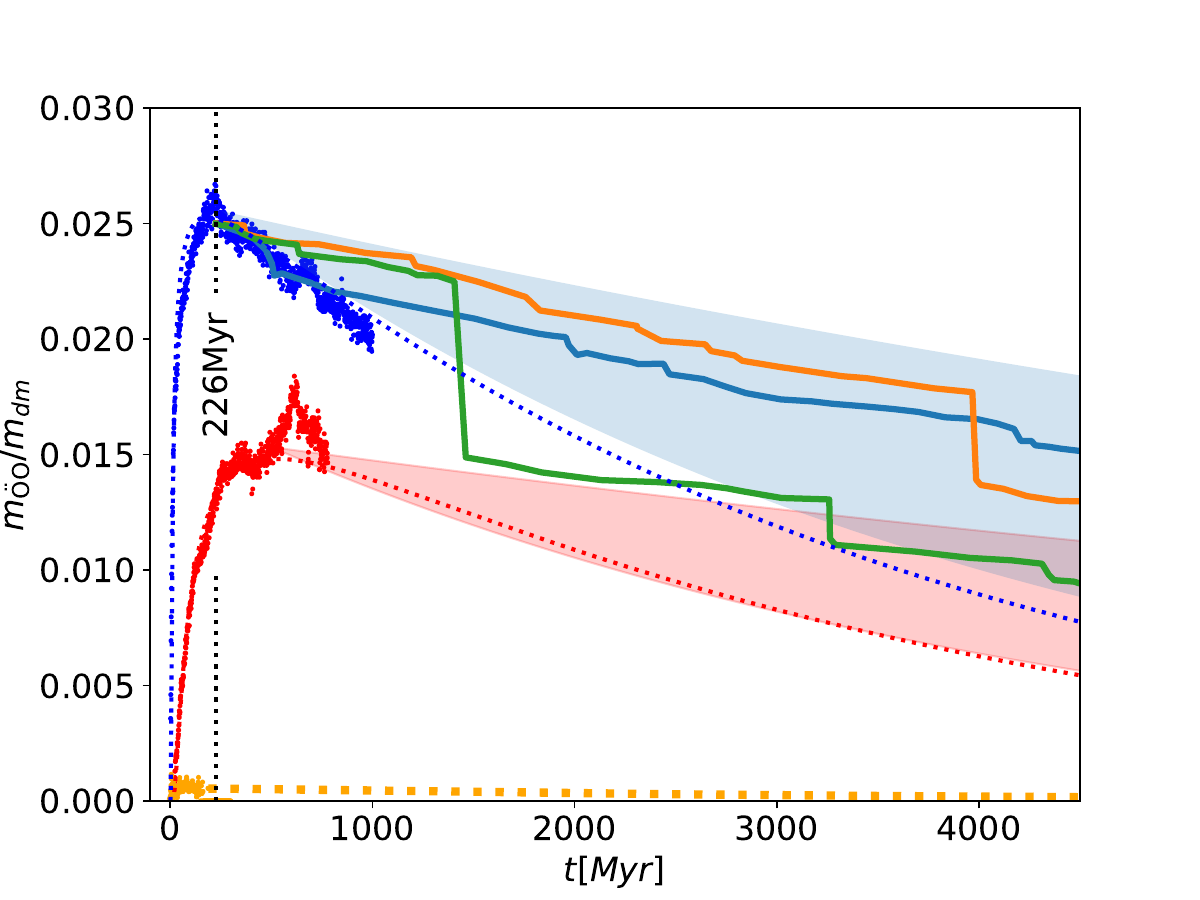}
\caption[]{Relative mass ($\mu_{\textrm{\"OO}} \equiv
  m_{\textrm{\"OO}}/m_{\rm dm}$) evolution of the \OOc. Blue dots (top
  left) result from simulating 1000\,Myr evolution of the Oort cloud
  for an isolated Solar system. Orange points (bottom) represent the
  formation of an \OOc\, in a stellar cluster; no appreciable Oort
  cloud forms in these models. The red dots give the mass evolution of
  the Oort cloud if the Sun left the parent cluster at an age of
  20\,Myr.
The dotted curves (blue, red, and orange) fit to the simulated data
using a fast-rise-exponential-decay (see \,\cref{Sect:Fred}), with
$t_{\rm rise} = 10$\,Myr for the blue and orange curves, $80$\,Myr for
the red curve, and all three curves have $t_{\rm decay} = 3500$\,Myr.
The solid curves give the \OOc\, mass-evolution for a star in a
Galactic orbit for three independent calculations by
\citep{2018MNRAS.473.5432H}. A similar mass evolution of the \OOc\, was
already calculated by \cite[][see their
  fig.\,7b]{2008Icar..197..221K}. The large jumps in these curves
result from encounters with other Galactic stars. The blue-shaded
region shows the uncertainty interval derived from these calculations
\citep{2018MNRAS.473.5432H}. The red-shaded region provides the
uncertainty interval of a similar analysis for a Solar system that
left the parent cluster after 20\,Myr. Both areas are bracketed by
$t_{\rm decay} = 4000$\,Myr and $t_{\rm decay} = 13\,000$\,Myr.
The vertical black-dotted curve indicates the moment of reaching the
maximum Oort-cloud mass for the isolated Solar system (see \cref{Sect:Methods}); from
this moment, Galactic erosion dominates the Oort cloud's evaporation.
\label{fig:OCGalacticOrbit}
}
\end{figure}

In \cref{fig:OCGalacticOrbit}, we present the mass-evolution of the
\OOc\, in the simulated Solar system. Initially the \OOc\, grows
exponentially on a timescale of $t_{\rm rise} =
9.3^{+7.6}_{-4.4}$\,Myr. The maximum relative \OOc\, mass of
$\mu_{\textrm{\"OO}} = 0.027\pm0.006$ is reached at $t_{\rm max} = 226
\pm 24$\,Myr (see~\cref{fig:OCpeakmass}, and consistent with
earlier calculations \citep{2008Icar..196..274B}). By that time
$62\pm2$\,\% of the ejected planetesimals are lost from the Solar
system, consistent with earlier \OOc\, formation efficiency estimates
\citep{1979MNRAS.187..445Y}. The lost planetesimals become free
floaters in the Galactic disk, such as 'Oumuamua
\citep{2017Natur.552..378M}.

After reaching the maximum mass, the \OOc\, erodes by losing comets
through tidal stripping by the Galactic potential and passing stars
\cite[see also ][]{2008Icar..197..221K}. Our calculations lasted for a
Gyr, during which the half-life of the \OOc\,$t_{\rm decay} \sim
3500$\,Myr to $5000$\,Myr, consistent with \citep{1985AJ.....90.1548H}
\cite[][find a slightly longer decay time of $t_{\rm decay} =
  4000$\,Myr to $13\,000$\,Myr, see also \cref{fig:OCGalacticOrbit}
  and \cref{fig:OCmassevolution} in the
  appendix]{2018MNRAS.473.5432H}. If the Solar system has evolved in
isolation, the \OOc\, should then have been roughly $1.9$ to $2.2$
times as massive at its maximum mass today, or between $0.4$ and
$4.5$\,\MEarth\, at the peak. With a retention efficiency of $\sim
0.027$, and accounting for erosion, the initial disk must have
contained $m_{\rm dm} \simeq 15$ to $170$\,\MEarth\, in solids after
accounting for the planets' masses.  This range is consistent with the
estimated $\sim 20$\,\MEarth\, based on the eccentricity dampening of
the giant planets \citep{2012AJ....144..117N}. The circum-stellar disk
can then at most have lost $\sim 5$\,\MEarth\, before the Oort cloud
started forming, resulting in an effective formation efficiency of
$\mu_{\textrm{\"OO}} \sim 0.02$, only slightly less than our estimate
from isolated evolution, see \cref{fig:OCGalacticOrbit}. The Sun then
cannot have lost much \OOc\, material while it was a cluster member,
before the tidal field manages to lift their perihelia.

\begin{figure}
\center
\includegraphics[width=\columnwidth]{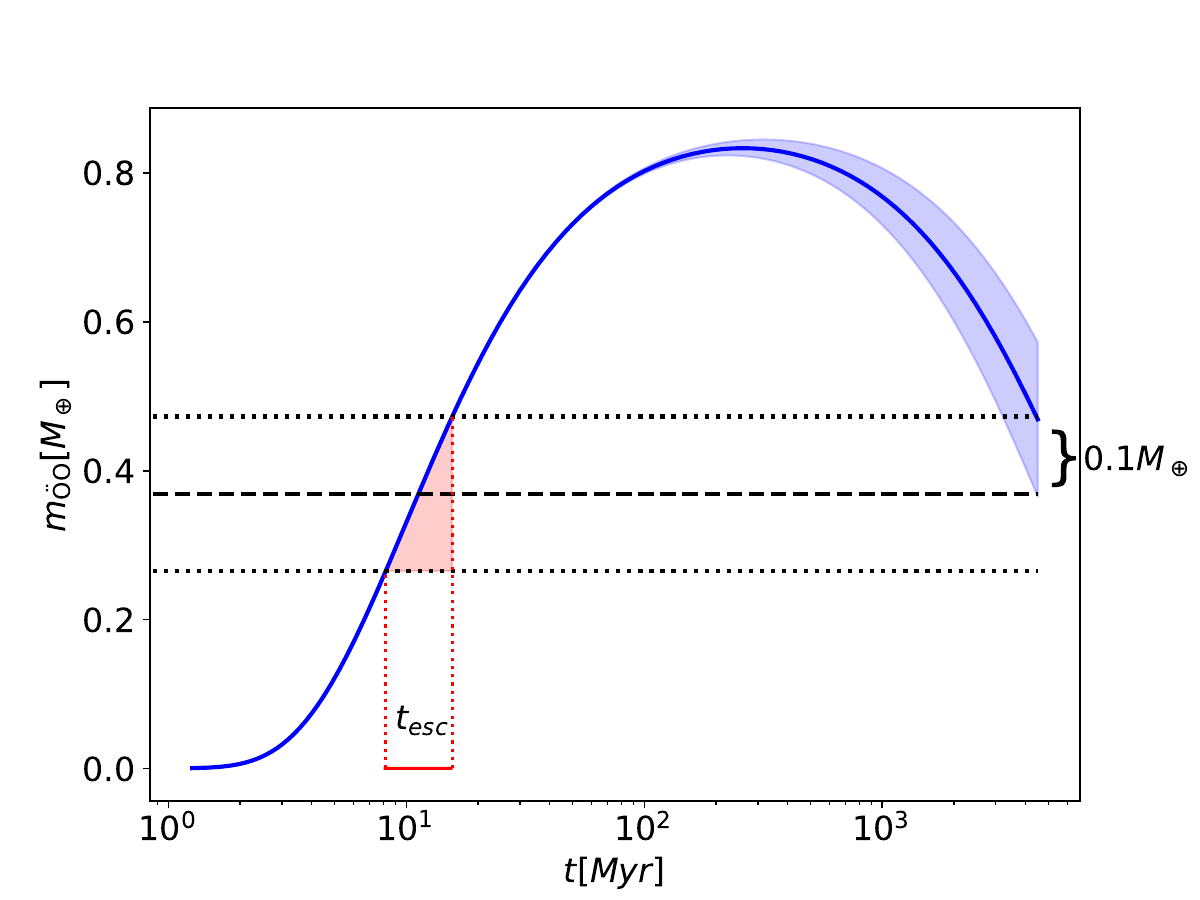}
\caption[]{Reconstruction of the time when the Sun left the parent
cluster for a specific set of parameters. The dark blue curve shows
the \OOc's mass-evolution, assuming it was born in isolation
(fast-rise-exponential-decay function with a maximum mass of
$\mu_{\textrm{\"OO}} = 0.027$, $t_{\rm rise} = 10$\,Myr, and $t_{\rm
decay} = 7500$\,Myr, see~\cref{Eq:FRED}). The blue-shaded area
gives the uncertainty of the \OOc\, for $t_{\rm decay} = 5000$\,Myr
to $10\,000$\,Myr (see also \cref{fig:currentOCmass} in the main
paper). The dashed line indicates $0.38$\,\MEarth\, below the
endpoint of the solid blue curve. The intersections between the
dashed line and the solid blue curve (to the left) indicate the
moment the Sun left the parent cluster at $t_{\rm esc} =
12\pm4$\,Myr.
The horizontal dotted lines are identical to the dashed line, but
present the extremes for the blue-shaded region; they define the
uncertainty in $t_{\rm esc}$, indicated in red.
\label{fig:OCmassevolution}
}
\end{figure}

\section{Oort cloud survival in a perturbing environment}

Irrespective of the birth cluster's mass, the \OOc\, cannot form while
the Sun was a member because its formation process is susceptible to
infinitesimal perturbations, and the cluster density within its Hill
sphere in the Galactic potential is independent of mass. From
adiabatic perturbations \citep{1985AJ.....90.1548H} derives an
exponential decay time-scale of $t_{\rm decay} \simeq 2100 (v_{\rm
  km/s}/a_{\rm 10^4au}^3)\, {\rm Myr}$. A planetesimal in an $e \sim
0.999$ orbit with a semi-major axis of $a = 55\,000$\,au at apocenter
would effectively co-move with the Sun. With a typical cluster
velocity-dispersion of $v \sim 1$\,km/s such a planetesimal escapes on
a timescale of $\sim 1.6$\,Myr \cite[see Eq. 47
  of][]{1985AJ.....90.1548H}, almost an order of magnitude shorter
than its orbital period. The formation of the \OOc\, in the early
Solar system must have been a race against the clock. As planetesimals
are ejected by repeated encounters with the giant planets near
pericenter (with the threat to become unbound), they gradually reach
further out into the \OOc\, to be kicked from their orbit by passing
cluster members. We simulate Solar system evolution in a clustered
environment to quantify this process. The results are summarized in
\cref{fig:OCGalacticOrbit}.

\begin{figure}
\center
\includegraphics[width=\columnwidth]{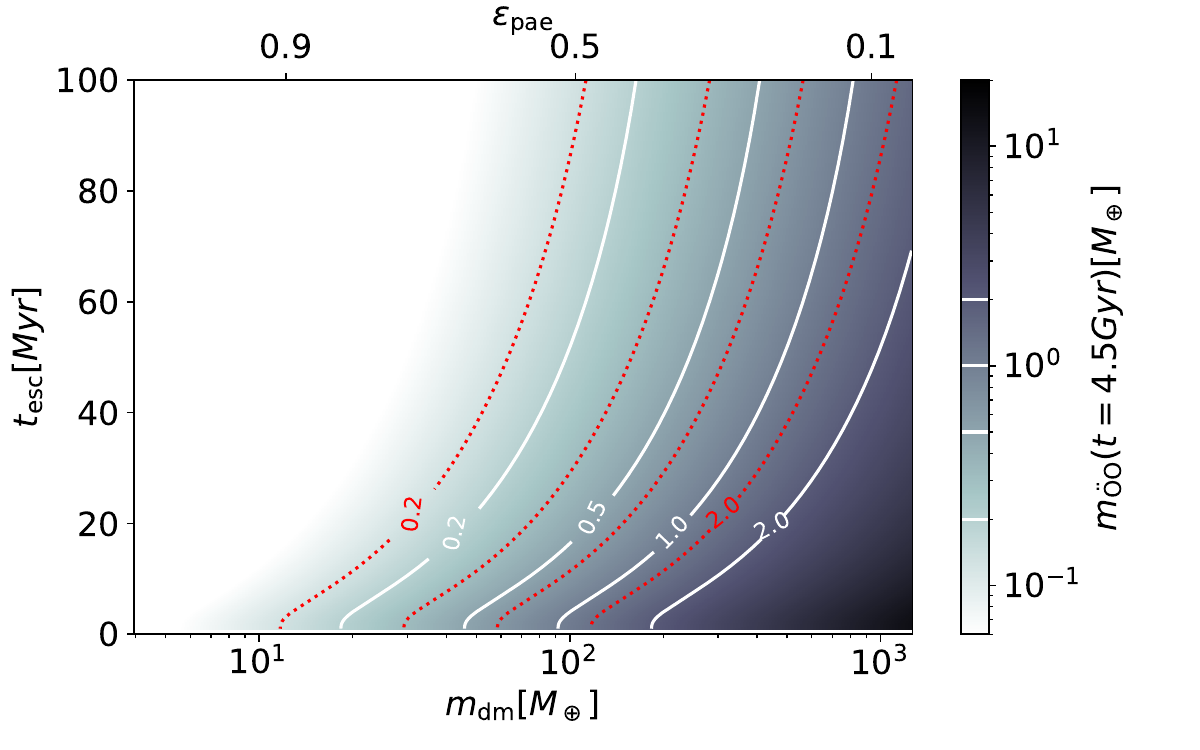}
\caption[]{Today's Oort cloud mass ($m_{\textrm{\"OO}}$, shades) as a
  function of the planetesimal disk-mass ($m_{\rm dm}$) and the Sun's
  escape time $t_{\rm esc}$. Here $m_{\rm dm}$ is the leftover mass of
  the debris disk after the giant planets' cores formed (here the mass
  in planets' cores is $m_{\rm pc}$). Along the top axis we express
  $m_{\rm dm}$ in terms of the planet-formation efficiency
  $\epsilon_{\rm pae} \equiv m_{\rm dm}/(m_{\rm dm}+m_{\rm pc})$. The
  curves indicate the current mass of the \OOc\, (in units of \MEarth)
  for a decay time scale $t_{\rm decay} = 5000$\,Myr (white) and
  $t_{\rm decay} = 10000$\,Myr (red), see also the color bar (to the
  right). Here we calculated $m_{\textrm{\"OO}}$ from the analytic
  model \cref{Eq:FRED} presented in ~\cref{Sect:Fred} (based on
  \cref{fig:OCGalacticOrbit}).
\label{fig:currentOCmass}
}
\end{figure}

The current masses of the Oort cloud and the detached populations
(outside the gravitational influence of Neptune but within $\sim
1000$\,au) are uncertain, but both depend on the moment the Sun leaves
the parent cluster. The normalization in our calculations depends on
the total mass of planetesimals in the circum-stellar disk after the
giant planets' cores formed. In \cref{fig:currentOCmass}, we present
the relation between initial disk mass, escape time ($t_{\rm esc}$)
and today's \OOc\, mass. For $m_{\rm dm} \aplt 30$\,\MEarth\, (or a
planet accretion efficiency of $\epsilon_\textrm{pae} \equiv m_{\rm
  dm}/(m_{\rm dm}+m_{\rm pc}) \apgt 0.77$) an \OOc\, of
$m_{\textrm{\"OO}} \apgt 0.5$\,\MEarth\, can form if the the Sun left
the cluster in $t_{\rm esc} \aplt 20$\,Myr. A more massive \OOc\, can
only form if the Sun left even earlier, or if the initial disk mass
was considerably higher \cite[which contradicts the constraints
  in][]{2012AJ....144..117N}.

The timescale of reaching the maximum \OOc\, mass is rather
insensitive to the cluster mass. However, The Oort-cloud formation
efficiency depends on the growth rate of the giant planets, their
orbital migration, and the extent of the primordial solar nebula
\citep{2007Icar..191..413B}.  If migration happens early
\citep{2022Natur.604..643L,2024A&A...688A.202G}, an Oort cloud can
only form when the sun leaves the cluster within $\sim 20$\,Myr after
migration ends.  But if migration time scale exceeds $\sim 20$\,Myr
\citep{RIBEIRO2020113605}, the Sun should have left before the
migration ends in order to grow an Oort cloud.

The processes in the star cluster that prevent the formation of the
\OOc\, stimulate the growth of the \HOc. Perturbations due to nearby
stars and the cluster potential tend to circularize their orbits,
detaching them from the planets' influence \citep{2015MNRAS.451..144P}.
In the cluster, the mass of the \HOc\, grows on a timescale of $t_{\rm
  rise} \sim 100$\,Myr to a maximum of $\mu_{\textrm{OHc}} \sim
0.048$. The mass of the detached population is hardly affected by the
cluster, and the Galactic tidal field has a negligible effect on the
long-term evolution of the detached population or on the \HOc; up to
$\sim 10$\,\% of the \OOc\, can be replenished by migration from the
\HOc\, due to passing stars and giant molecular clouds
\citep{2006Icar..184...59B,2008Icar..196..274B}.

\begin{figure}
\center
\includegraphics[width=\columnwidth]{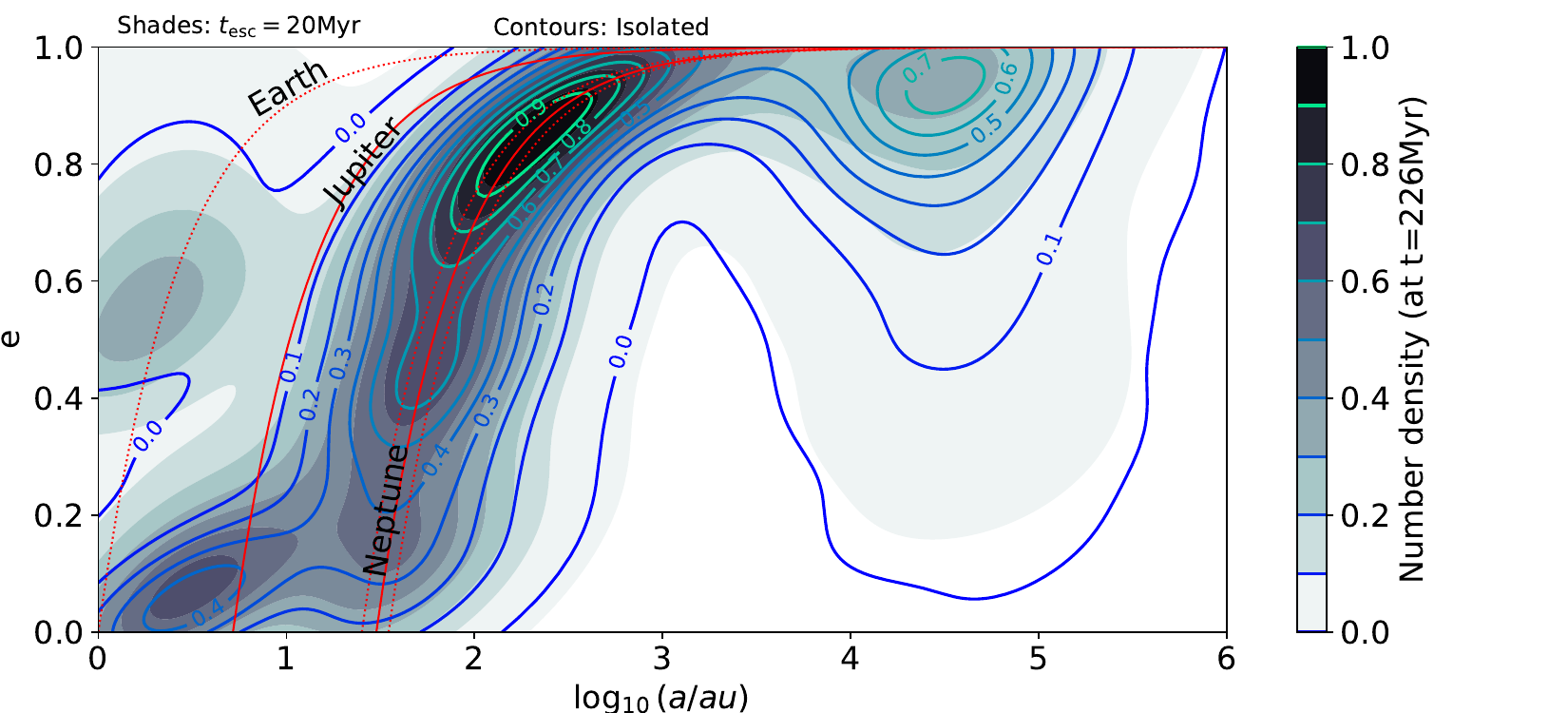}
\includegraphics[width=\columnwidth]{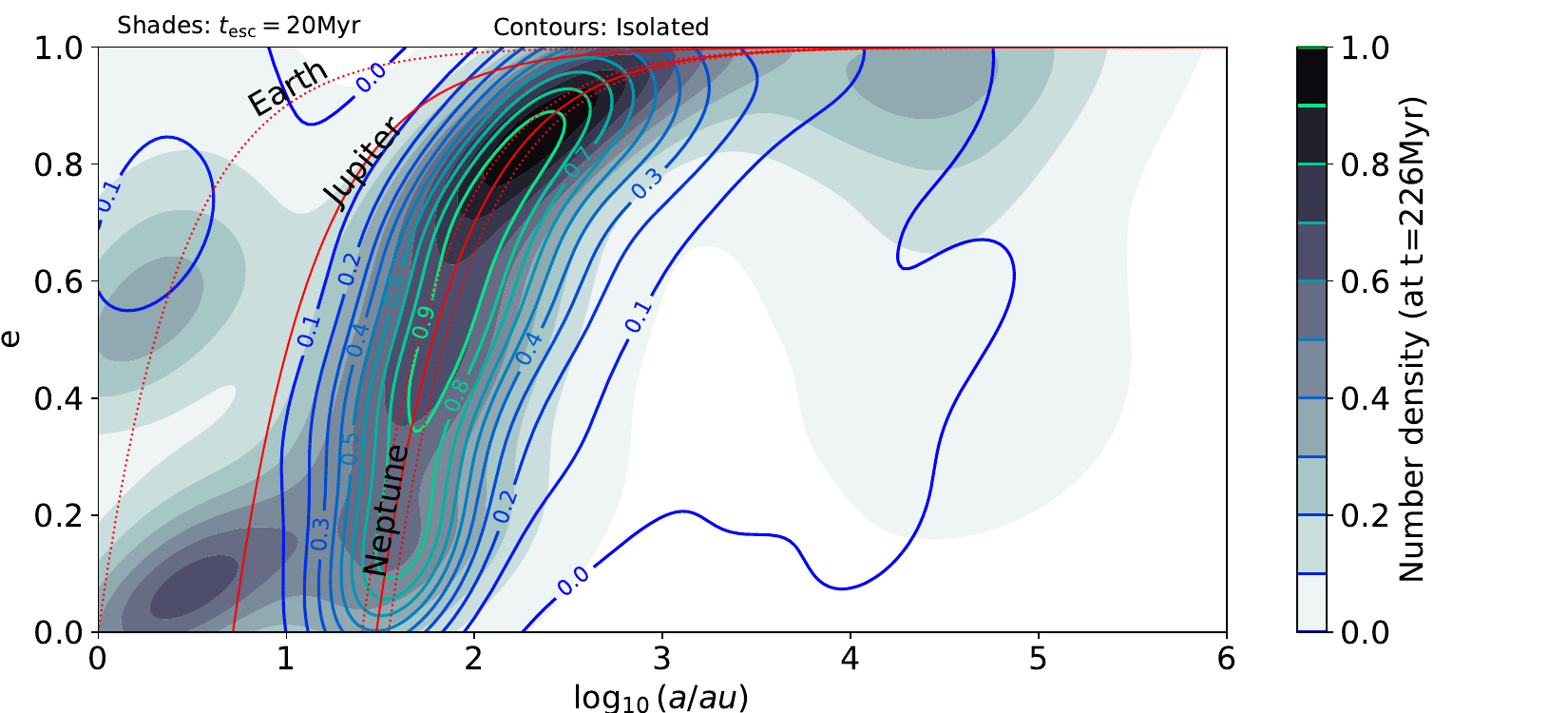}
\caption[]{Distribution of semi-major axis and eccentricity of the
  Solar system planetesimals. The distributions are presented at their
  peak, at the age of 226\,Myr. The gray shades give the probability
  density function (linearly interpolated) for the Solar system that
  was ejected 20\,Myr after the formation of the giant planets.
The red curves indicate the influence of Earth (left dotted curve for
reference), Jupiter (middle), and Neptune (rightmost curve). The two
dotted curves around Neptune's influence indicate the range within 5
Hill radii of Neptune. Top panel: The contours (same interpolation)
give the probability density distribution for the Solar system that
was isolated for its entire lifetime.\\
Bottom panel: The contours (same interpolation and range) give the
probability density distribution for the Solar system that remained in
the cluster for its entire lifetime.
\label{fig:S2_ae_at_226Myr}
}
\end{figure}

In \cref{fig:S2_ae_at_226Myr} we present distributions in semi-major
axis and eccentricity for a hybrid simulation in which we evolved the
Solar system for 20\,Myr in a stellar cluster and continue the
calculation in isolation (see also \cref{fig:OCGalacticOrbit}). We
overlay the planetary system that evolved in isolation (contours in
top panel), and the one that remained in the cluster (bottom
panel). The distributions show similar morphology for the detached
population around Neptune's influence (the rightmost red curve), but
very different distributions around the Oort-Hills and \OOc s. Since
the \HOc\, preserves signatures of the Sun's birth environment
\citep{2016MNRAS.457.4218J} the discovery of planetesimals in $500$\,au
to $\sim 5000$\,au orbits with an eccentricity of $e=0.2$ to $0.9$
would put interesting constraints on the birth cluster parameters and
the Sun's encounter history.

A prolonged evolution in isolation or in a clustered environment also
has consequences for near-Earth objects (NEOs). The number of NEOs at
the moment the Oort cloud reaches its peak mass is typically 5 times
higher in the models where the Sun escapes in 20\,Myr compared to the
Sun born in isolation. Such a higher local planetesimal density, as
visible in \cref{fig:S2_ae_at_226Myr}, may have interesting
consequences for the earliest cratering record of the inner planets
and Moon \cite[see also][]{2024NatAs...8.1380P,2024ApJ...972L..21P}.

\section{Discussion}

According to \citep{2018MNRAS.473.5432H,2019MNRAS.490...21H}, the Sun
accreted the Oort cloud from other stars.  Such accretion can only be
successful if other stars also have their own Oort clouds. However,
since no \OOc\, forms while a star is a cluster member, material
exchange among stars is prevented. Some inner material with $a \aplt
10\,000$\, au and low eccentricities (e $\aplt 0.9$) can be exchanged
\citep{2016MNRAS.457.4218J,2024ApJ...972L..21P}, but this material is
easily lost again when accreted in wide $a \apgt 10\,000$\,au
orbits. The \HOc~\citep{2017A&A...599A..91B,2019MNRAS.490...21H} and
the detached population \citep{2015MNRAS.453.3157J}, on the other hand,
are quite likely to be rich in extra-solar planetesimals due to
exchange of material during close encounters in the young cluster.

The recently discovered population of 92 planetesimals with $a>50$\,au
(two of which with $a\apgt 10^3$\,au), found with the Horizon mission
\citep{2024PSJ.....5..227F}, have eccentricity $e = 0.61 \pm 0.16$ and
inclination $i = 13.1^\circ \pm 15.4^\circ$; Their orbits are
inconsistent with a formation in situ and from being ejected by
giant-planet migration \citep{2020Icar..33913605D}, both of which
predict planetesimals beyond 50\,au to have low eccentricities and low
inclinations in a narrow range.  Their orbits, however, are consistent
with external perturbations induced by passing stars which naturally
lead to a wide range in eccentricity and inclination, and with much
higher values
\citep{2016MNRAS.457.4218J,2019MNRAS.490...21H,2024ApJ...972L..21P}.

Alternatively, part of the Oort cloud could be captured from
free-floating debris in the parent cluster \citep{2010Sci...329..187L},
or from freezing out planetesimals into bound pairs with the Sun,
much in the same way wide binaries form \citep{2011MNRAS.415.1179M} or
planets are captured \citep{2012ApJ...750...83P}. The rate calculated
through gas expulsion in the young cluster by
\citep{2010Sci...329..187L} overestimates the efficiency of this
process because of their assumed sudden (in $10^4$\,yr) removal of the
primordial gas from the cluster central region. Recent calculations of
cluster formation indicate that gas is ejected in much smaller
quantities, slower, and from the cluster periphery rather than from
the center \citep{2021MNRAS.502.3646G,2024A&A...690A.207P}, in contrast
to the assumptions adopted in \citep{2010Sci...329..187L}. The
remaining capture rate in a cluster of $N$ stars derived by
\citep{2011MNRAS.415.1179M,2023ApJ...955..134R} gives only a small
fraction of 0.03 to 0.06, and $\propto 1/N$. Although the Sun and
potentially each of the other $N \sim 3\,000$ to $10^4$ stars in the
cluster have already lost a considerable portion ($\sim 60$\%) of
their planetesimals by the time the Sun escapes the cluster, the
fraction of objects remaining bound to the escaping Sun would be only
$10^{-4}$ to $10^{-5}$. This is substantially smaller than the
fraction of native planetesimals that populate the Oort Cloud ($\sim
0.027$).

\section{Conclusions}

The existence of a rich \OOc\, can then only be reconciled with an
early escape from the parent cluster, much earlier than expected based
on the estimated lifetime of a virialized birth cluster, which easily
exceeds 100\,Myr \citep{2010ARA&A..48..431P}. If the Sun was ejected
from the young cluster by a strong encounter, the \OOc\, can only have
formed after this event, and the majority of material must be native
to the Solar system. The whole cluster likely dissolved
quickly. Roughly half to 90 percent of the stars are born in
non-virial or fractal-structured environments
\citep{2022MNRAS.515.2266A} that experience violent relaxation in the
first 10\,Myr \citep{2006MNRAS.373..752G}. Such a cluster dissolves
shortly thereafter \citep{2009Ap&SS.324..259G}, consistent with our
estimated time spent in the clustered environment. For the Sun, we
then favor a dense (half-mass density $\apgt 10^3$stars/pc$^3$)
non-virial birth environment.

It is hard to derive a minimum amount of time the Sun has spent in the
cluster, but it must have been a member long enough to experience a
strong encounter, if only to truncate the circumstellar disk at the
observed Kuiper cliff \cite[at $\sim
  50$\,au,][]{2024MNRAS.527L.110D,2024NatAs...8.1380P}, and to explain
the hot trans-Neptunian population
\citep{2014MNRAS.444.2808P,2024PSJ.....5..227F}, or the irregular moons
\citep{2024ApJ...972L..21P}.  In that case the $\apgt 50$\,au
planetesimals were scattered in the last encounter to their current
orbits, either originating from the Sun's disk, or being captured from
the encountering star's disk.  An early escape also has consequences
for the expected number and the proximity of supernovae in the infant
Sun's neighborhood. The first supernova typically happens between
$8$\, and $10$\,Myr after the cluster's birth
\citep{2019A&A...622A..69P}. Such a nearby supernova could explain the
solar system's abundance in short-lived radionuclides
\citep{2023A&A...670A.105A}, and the observed $\sim 7^\circ$ tilt in
the ecliptic to the Sun's equatorial plane \citep{2019A&A...622A..69P}.
We expect that the birth cluster was non-virial, probably with more
stars than the hitherto estimated $2500\pm 500$ stars
\citep{2009ApJ...696L..13P,2010ARA&A..48...47A,2023A&A...670A.105A},
and we left the nest within $20$\,Myr after the giant planets formed.

\section{Acknowledgements}

We thank Anthony Brown, Steven Rieder and Eiichiro Kokubo for
discussions. Many of our calculations were performed on the two
workstations of Ignas Snellen and Yamila Miguel; we thank their
students for their patience.

This work used the Dutch national e-infrastructure and in particular
the Cartesius supercomputer with the support of the SURF Cooperative
using grant numbers EINF-12718 and 2024.058.

\subsection{Energy consumption of this calculation}
Considering the climate and energy consumption, we desire to report on
the damage our work has done to the environment. The calculations
have been running for about four months on dual 128-core Xeon-equipped
workstations, totaling about 1024 core-days. Assuming a CPU
consumption of $12$ Watt hr$^{-1}$, the total energy is roughly
$0.3$\,MWh. For an emission intensity of $0.283$ kWh kg$^{-1}$
\citep{doi:10.1002/cpe.3489}, our calculations emitted $1$\,tonne CO2.

\subsection{Data/Code Availability}
This work is carried out using the Astronomical Multipurpose Software
Environment (AMUSE). AMUSE is open source under the Apache-2.0
license, and the entire source code can be downloaded via
\url{https://github.com/amusecode/amuse}. The simulations in this work
were performed using AMUSE version v2024.6.0).

We used the LonelyPlanets AMUSE script, which is open source under the
Apache-2.0 license, and the entire source code can be downloaded via
\url{https://github.com/spzwart/LonelyPlanets}

The simulation data, and specific scripts to generate the figures in
this manuscript can be downloaded through FigShare at
\url{https://figshare.com/account/items/25744584} (DOI:
10.21203/rs.3.rs-4257777/v1).





\setcounter{section}{0}

\begin{appendix}
  

\section{Isolated planetary systems}\label{Sect:SM:isolation}\label{Sect:Methods}

We simulate isolated planetary systems and those perturbed in a
clustered environment using the AMUSE framework. AMUSE
\citep{2018araa.book.....P} is a high-performance software instrument
for simulating four fundamental physics domains: gravity,
hydrodynamics, stellar evolution, and radiative transfer. It can be
used for a wide variety of applications. The various codes that solve
for these domains are interchangeable and can run concurrently on
different resolutions and scales.

For simulating an isolated planetary system, we start with the Sun and
the four giant planets using ephemerides at Julian date 2588106.62749,
with their current masses. We subsequently added a population of minor
bodies in the ecliptic with circular ($e \lteq 0.01$) orbits between
1\,au and 42\,au around the Sun
\citep{1988Sci...242..547M,2024A&A...682A..43G}. The minor bodies have
no mass and no size. The surface density of the disk follows a $-1.5$
power-law and has a Toomre Q-parameter of 1.0
\citep{1972ApJ...178..623T}. We generate planetary systems using the
AMUSE routine {\tt ProtoPlanetaryDisk}. We rotate the Solar system
(planets and planetesimals) along the y-axis by $60.2^\circ$ and put
it in a circular orbit in the Galactic plane at a distance of 8.5\,kpc
along the x-axis. We then move the center of mass of the simulated
Solar system by 7.5\,pc in the Galactic z-coordinate, and give it an
additional velocity of 10.1 km/s towards the Galactic center. This
places the Solar system in a slightly elliptical and inclined orbit in
the Galactic potential, with the Solar system's ecliptic inclined by
about $60^\circ$. We adopt the simple Galactic potential presented in
\citep{2018araa.book.....P}.

The planetesimals in our calculations are massless. Still, we can
consider our results in light of the planet formation efficiency. For
this estimate, we focus on the mass in rocky cores of the giant
planets. Jupiter's core mass has an upper limit of 45\,\MEarth\,
\citep{2017GeoRL..44.4649W}, for Saturn we adopted 25\,\MEarth\,
\citep{2024A&A...682A..43G}, 13\,\MEarth\, for Uranus and
15.5\,\MEarth\, for Neptune
\citep{2007ApJ...671..878D,2011ApJ...726...15H}. The upper limit to the
mass in a giant planet core is then $m_{\rm pc} \simeq
98.5$\,\MEarth. The original mass of the disk in solids, before the
planets formed must then have been at least $98.5$\,\MEarth\, plus the
current mass of minor planets and planetesimals give a minimum disk
mass of $\sim 100$\,\MEarth.  The mass scaling from test particles to
appropriate mass units is realized by adopting an initial disk mass,
such as presented in \cref{fig:currentOCmass}.

The dynamical evolution of the planetary systems is performed with the
direct N-body code Huayno, using the drift-kick-drift integrator with
time-step parameter $\eta = 0.03$. The dynamics of the Solar system
and the Galaxy are coupled using the bridge method
\citep{2007PASJ...59.1095F} with a time-step of 0.01\,Myr. This is a
conservative choice considering the static nature of the background
potential, but it allows us to use the same time step when including
stellar encounters, which require a high time resolution.  A bridge
time step of 0.01Myr corresponds to the period of a circular orbit at
about 460au, which is more than a factor 60 shorted than inner edge of
the \OOc\, boundary at $30\,000$\,au.

\section{Oort cloud formation in isolation}\label{Sect:SM:I:Oort}

We run 27 simulations with $4$ planets and $10^3$ planetesimals. In
principle, we could have performed a single simulation with 27\,000
planetesimals at the same cost, but we now can address the
stochasticity in the results due to the chaotic dynamics of the
planetary system, and we adopt the same planetary systems in a
dynamical environment, as we discuss further down.

For each simulation we identify various dynamical classes. Rather
than listing the entire procedure for identifying minor bodies, we
provide the selection criteria for the various Oort-cloud objects and
the remaining trans-Neptunians (which includes the various Kuiper-belt
families).

The orbital elements are calculated using a Kepler solver in the Solar
system's barycenter. The procedure for identifying dynamical classes
is as follows:
\begin{enumerate}
\item First, classify the bound planetesimals (with $e < 1$).
\item Then we select those with pericenter distance $a(1-e)>5$\,Hill
radii further away from the Sun than the planet Neptune.
\begin{enumerate}
\item Those with $a \gteq 30\,000$\,au populate the \OOc.
\item The remaining planetesimals with $a \gteq 1\,000$\,au populate the \HOc.
\item the left over (detached but with $a < 1\,000$\,au) are the
detached objects.
\end{enumerate}
\end{enumerate}

We adopt an inner edge of the \OOc\, of $r_{\rm inner} = 30\,000$\,au,
but investigating \cref{fig:S2_ae_at_226Myr} of the main paper, we
could consider an inner limit around 10\,000\,au to be more
distinctive, as this is also around the region beyond which the
Galactic tidal field starts to become important.

\begin{figure}
\center
\includegraphics[width=\columnwidth]{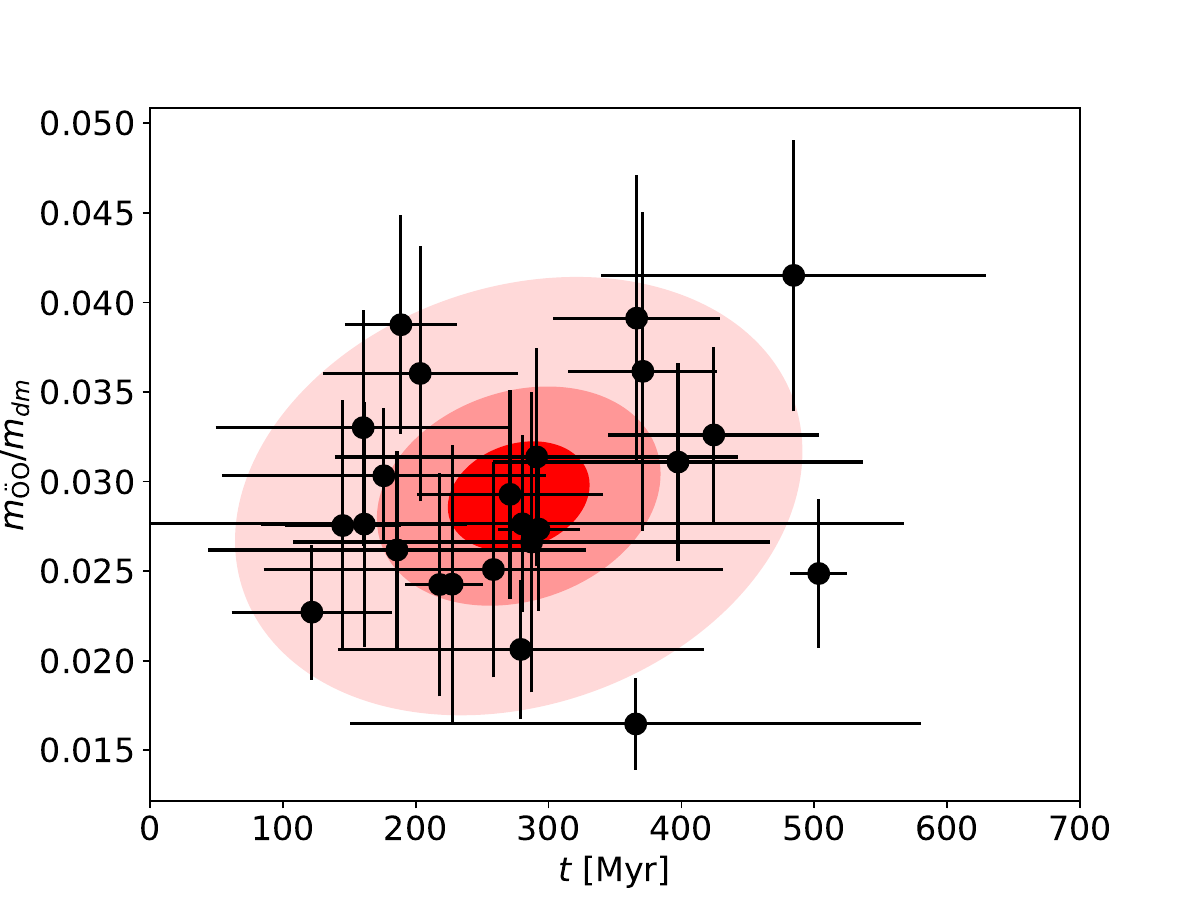}
\caption[]{Peak in the relative mass of the \OOc\, ($\mu_{\textrm{\"OO}}$)
for the isolated Solar system. The black bullet points represent
the measured maximum mass of the \OOc, with 1 standard deviation
uncertainty, measured at the time the maximum Oort cloud mass is
reached. The shaded areas show the uncertainty obtained from 27
calculations centered on the global mean with the error ellipse for
0.5 (darkest shade red), 1.0, and 2.0 (lightest shade) standard
deviations. A least-squares fit of a linear function to the data
does not give a significant result. The geometric mean of the 27
bullet points is at a mass of $m_{\rm dm} = 0.027\pm0.006$, at an
age of $226 \pm 24$\,Myr.
\label{fig:OCpeakmass}
}
\end{figure}

Once classified, we can describe the evolution of the various detached
populations. We start by measuring the time and number at the peak of
the \OOc\, in each of our simulations. The results are presented in
\cref{fig:OCpeakmass}, showing the time and maximum of the Oort cloud
in each of the 27 simulation and the statistical average.

\section{Numerical noise and chaos}\label{Sect:SM:I:chaos}

In principle, the planets in each of our isolated systems should
evolve identically because they start with identical initial
realization for the Sun and planets, whereas the planetesimals, which
are treated as test particles, have orbits randomly selected from the
initial conditions. Integrating these systems causes variations in the
round-off at the least significant digit. This round-off drives chaos
in our calculations, causing the orbital evolution of the planets to
be different for different planetary systems. To validate our results
in terms of its chaotic behavior of the system, and to ensure that the
results are not driven by the chaotic nature of the system, we
calculate the largest positive Lyapunov timescale $t_{\rm Ly}$ for
each of the 351 unique pairs of planetary systems.

The Lyapunov timescale is calculated by fitting the growth of the
difference between the normalized orbital semi-major axes of the four
planets. Fitting an exponential relation gives an estimate for the
relative growth of numerical errors due to round-off, and the Lyapunov
timescale is its reciprocal.

We find a mean Lyapunov timescale of $\langle t_{\rm Ly}\rangle =
800\pm230$\,Myr, with a tail to $t_{\rm Ly} = 4000$\,Myr containing
$\sim 20$\,\% of the cases. The Lyapunov timescale is of the order
of our calculation time ($1000$\,Myr), indicating that our
calculations are indeed chaotic, as expected, but this has not
affected the results. Upon inspection of the planets' orbital
parameters at the end of the simulations, we find slight variations of
the order of $\sim 0.4$\% in orbital separation. We consider these
small compared to the variations in orbital separation needed to
affect the results, which would be around an order of magnitude larger.

\section{Planetary system in a stellar cluster}\label{Sect:SM:cluster}

Planetary systems in a stellar cluster are simulated in two stages
using the LonelyPlanets AMUSE script. This new implementation of the
method, introduced in \citep{2019MNRAS.489.4311C}, is now independent
of the adopted N-body integrator, and it is available at
github\footnote{see https://github.com/spzwart/LonelyPlanets}. The new
LonelyPlanet script includes stellar evolution, collisions, and a
semi-analytic background Galactic potential.

Integrating a star cluster, including planetary systems, is
impractically expensive in terms of computer time. In LonelyPlanets,
we reduce these costs through a two-stage divide-and-conquer strategy
\citep{2019MNRAS.489.4311C,2020MNRAS.497.1807S}. In stage one, we
simulate a dynamical evolution and stellar evolution of a cluster of
$N$ stars in the Galactic potential: In stage two, we integrate the
planetary systems with only the nearest perturbing cluster
members. This strategy easily saves a factor of $10^4$ in computer
time.

In stage one, the stars' motion equations in the cluster are
integrated using an N-body code dedicated to cluster dynamics. For
this work, we opted for the 4th-order Hermite predictor-corrector
scheme \citep{1992PASJ...44..141M} Ph4 \citep{2018araa.book.....P}. The
coupling of the direct N-body code for the cluster dynamics and the
semi-analytic background potential of the Galaxy is realized using
bridge \citep{2007PASJ...59.1095F}. Bridge is a non-intrusive
hierarchical code-coupling strategy based on solving the combined
Hamiltonian as two separate parts using Strang splitting
\citep{doi:10.1137/0705041}. Both parts are later combined to form a
homogeneous and self-consistent solution using a kick-drift-kick
scheme, as was demonstrated in \citep{2020CNSNS..8505240P}.

\begin{figure}
\includegraphics[width=\linewidth]{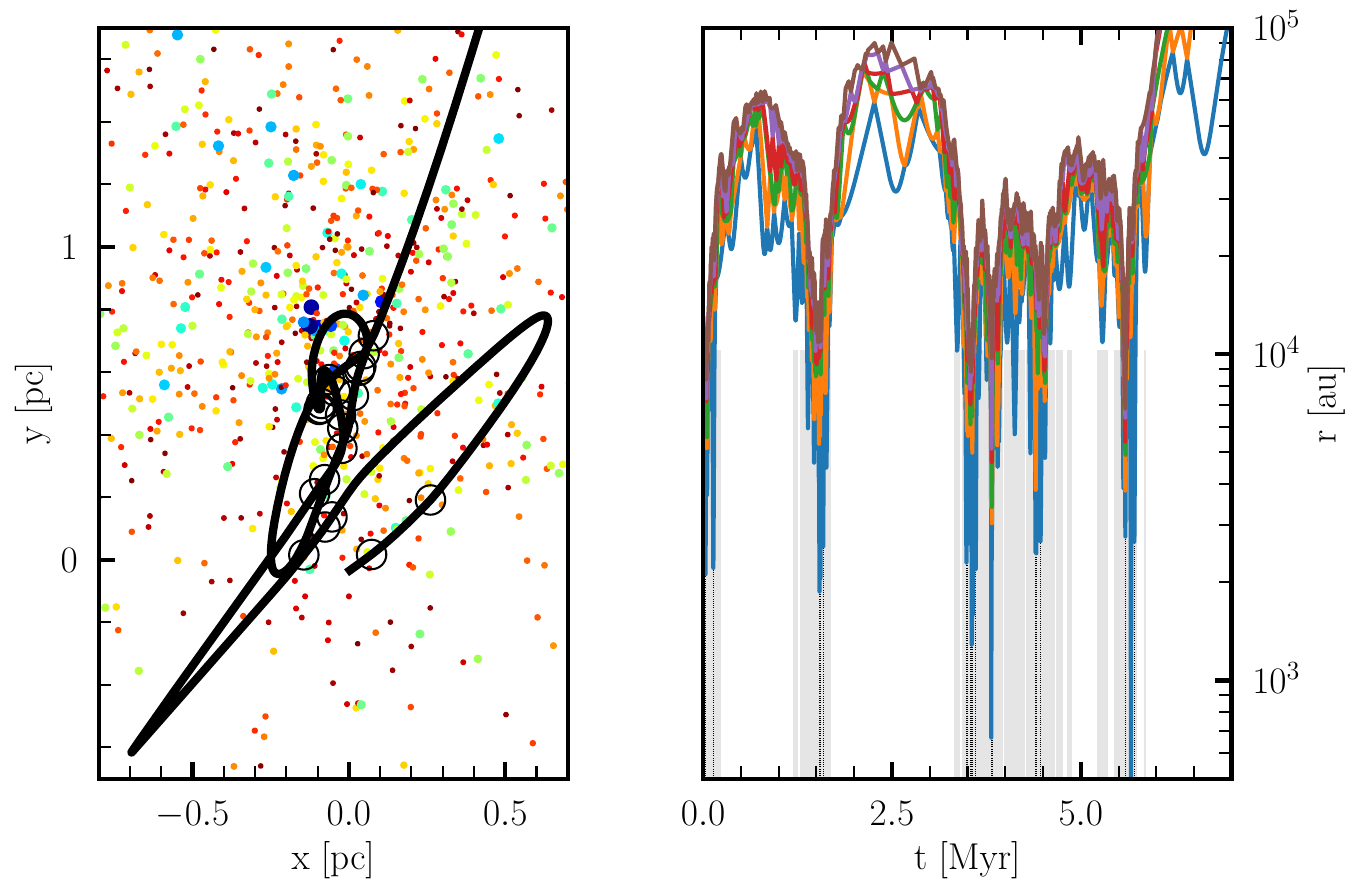}
\caption{ Initial cluster (left: rainbow colors and size represent
stellar mass), and the orbit of a randomly selected star evolved for
7\,Myr (black curve). Circles along the orbit indicate moments of
close approach. The right-hand panel shows the distance to the 6
nearest neighbors as the selected star orbits the cluster. The gray
shaded area indicates when another star approaches within
$10^4$\,au.
\label{fig:LonelyPlanets}
}
\end{figure}

While integrating the cluster, we keep track of the $N_{\rm cc}=6$
closest encountering stars and the strongest perturbers for a subset of
$27$ selected stars. We store the encounter information between
selected stars and perturbers at a fine time resolution of
$0.01$\,Myr. Stage one is illustrated in
fig.\,\ref{fig:LonelyPlanets} where we show the orbit of one selected
star in a cluster of $10^3$ stars over 7\,Myr. The right-hand panel
shows the distance for the selected star to its $N_{\rm cc} = 6$
nearest neighbors (which includes the 3 strongest perturbers). The
shaded regions in the right-hand panel indicate when the selected star
is strongly perturbed. In this particular case, the selected star
escapes the cluster after a particularly strong encounter at $\sim
5.7$\,Myr.

In stage two, each of the stars acquires a system of $N_{\rm p}=4$
planets and $N_{\rm a}=1000$ planetesimals. The planetesimals are test
particles; they do not feel each other's gravity and do not exert a
force on any of the planets or stars. We integrate the orbits of the
planets and planetesimals together with the earlier stored neighboring
stars, replenishing this population every 0.01\,Myr. In some sense, we
reconstruct the perturbed star's dynamical history while integrating it including its planetary system. Each planetary
system, including its perturbers, is integrated using the symplectic
coupled-components drift-kick-drift algorithm Huayno
\citep{2012NewA...17..711P,2014A&A...570A..20J}.

\begin{figure}
\includegraphics[width=\linewidth]{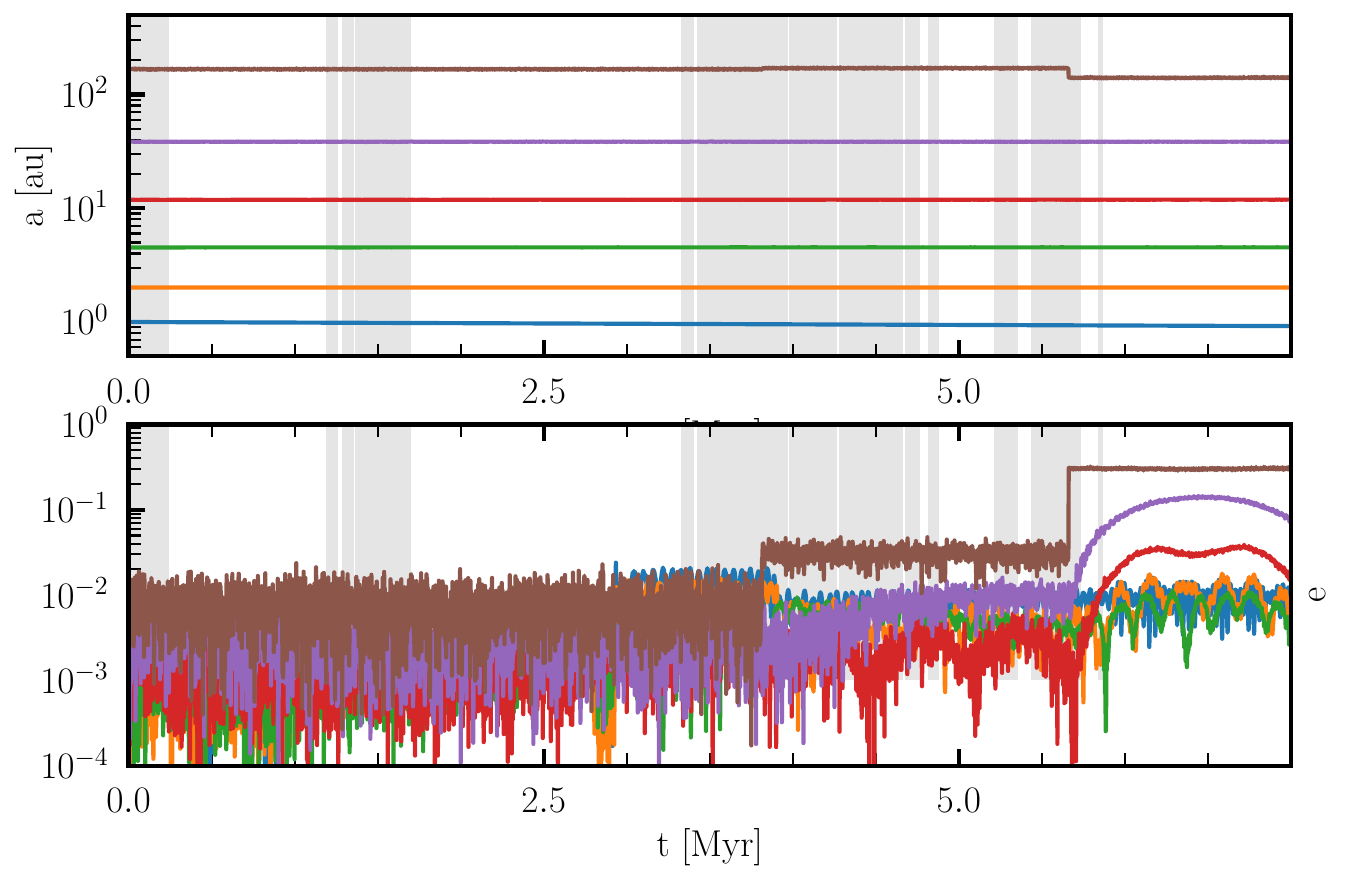}
\caption{Evolution of the orbital separation (top) and eccentricity
(bottom) for the selected star in fig.\,\ref{fig:LonelyPlanets}, but
now with 6 planets between 1\,au and 180\,au according to an
Oligarchic planet-formation model \citep{1998Icar..131..171K}. The
planetary system was integrated for 7\,Myr, while the 6 nearest
stars were taken into account in the gray-shaded areas. The initial
planetary system would have been stable if it evolved in isolation,
but the perturbations caused several planets to change their orbits
considerably. Most noticeable is the high increase in eccentricity
to 0.31 of the outermost planet due to the encounter that kicked the
star from the cluster (see~\cref{fig:LonelyPlanets}). An earlier
encounter at 3.8\,Myr already affected the outer planet's orbit.
\label{fig:LonelyPlanets_orbit}
}
\end{figure}

Stage two is illustrated in fig.\,\ref{fig:LonelyPlanets_orbit} where
we show the evolution of the orbits of 6 planets around the selected
star. The choice of the number of planets is arbitrary; it just shows
the workings of the code. The encounter that ejected the star from the
cluster (at $\sim 5.7$\,Myr) also caused a variation in the semi-major
axis and eccentricity in the outermost planets. This perturbation is
propagated to the inner planets' orbits on a secular timescale.

Integrating the planetary system requires ${\cal O}((1+N_{\rm p}+
N_{\rm cc})^2 + N_{\rm a})$ operations per time step. Once the star
escapes the cluster, its planetary system is unlikely to be perturbed
by nearby stars, but it remains affected by the Galactic tidal
field. The isolated planetary system is integrated by a symplectic
integrator, which bridges with the semi-analytic Galactic
potential.

Integrating the cluster of 1000 stars and 27 planetary systems for
1000\,Myr using a direct (symplectic) method would require some $1.7
\cdot 10^{21}$ flops (28108 stars, planets and planetesimals in total
integrated with a 1\,day time step, assuming 60 operations per force
calculation). Such a calculation would cost about two weeks on a
PetaFlop-scale computer. In LonelyPlanets the same calculations can be
done in only $6\cdot 10^{11}$ operations for the cluster simulation
and $1.7\cdot 10^{15}$ (about two weeks on a 128-core workstation)
operations for each of the planetary systems. The speedup due to
LonelyPlanets then amounts to a factor of $\apgt 10^4$ compared to
conventionally integrating the cluster with planets.

We performed several calculations of planetary systems in stellar
clusters. We adopted the same 27 planetary systems every time but
varied the cluster parameters.

We adopted virialized Plummer spheres with 100 to 6000 stars from a
Salpeter mass function between 0.1\,\MSun\, and 100\,\MSun\, with
virial radii of 0.5\,pc, 1.0\,pc and 1.5\,pc. These clusters cover a
wide range of initial half-mass densities from $3.7$\,pc$^{-3}$ to
$6000$\,pc$^{-3}$. We also performed simulations with a more complex
initial realization in which we adopted the results from a
hydrodynamics molecular cloud collapse simulation until the gas was
depleted
\citep{2023MNRAS.520.5331W,2024A&A...690A.207P,2024A&A...690A..94P}.

After initializing the cluster stars, we select 27 stars with a mass
closest to 1\,\MSun\, in each of these initial realizations and turn
them into 1\,\MSun\, stars. After this we correct for the mass change
by rescaling the system to virial equilibrium. Note that we did not
rescale the clusters resulting from the hydrodynamical simulation, as
it was initially not in virial equilibrium.

We continue each simulation to 1000\,Myr, but stop integrating each
Solar-equivalent system after losing a planet or if the number of
planetesimals drops below 50. Many of the planetary systems were
destroyed before that time, particularly in those runs with a high
stellar density. Eventually, we analyze the data for the surviving
planetary systems.

The initial stellar distribution did not make a statistically
significant difference for the surviving planetesimal orbital
distributions in the \OOc. The resulting mass evolution of the \OOc,
is summarized in the main paper \cref{fig:OCGalacticOrbit}. In all the simulations,
the formation process of an \OOc\, turns out to be too fragile to
survive a long exposure in a clustered environment.

\section{Further analysis on the growth and erosion of an Oort cloud}\label{Sect:Fred}

\begin{figure}
\center
\includegraphics[width=\columnwidth]{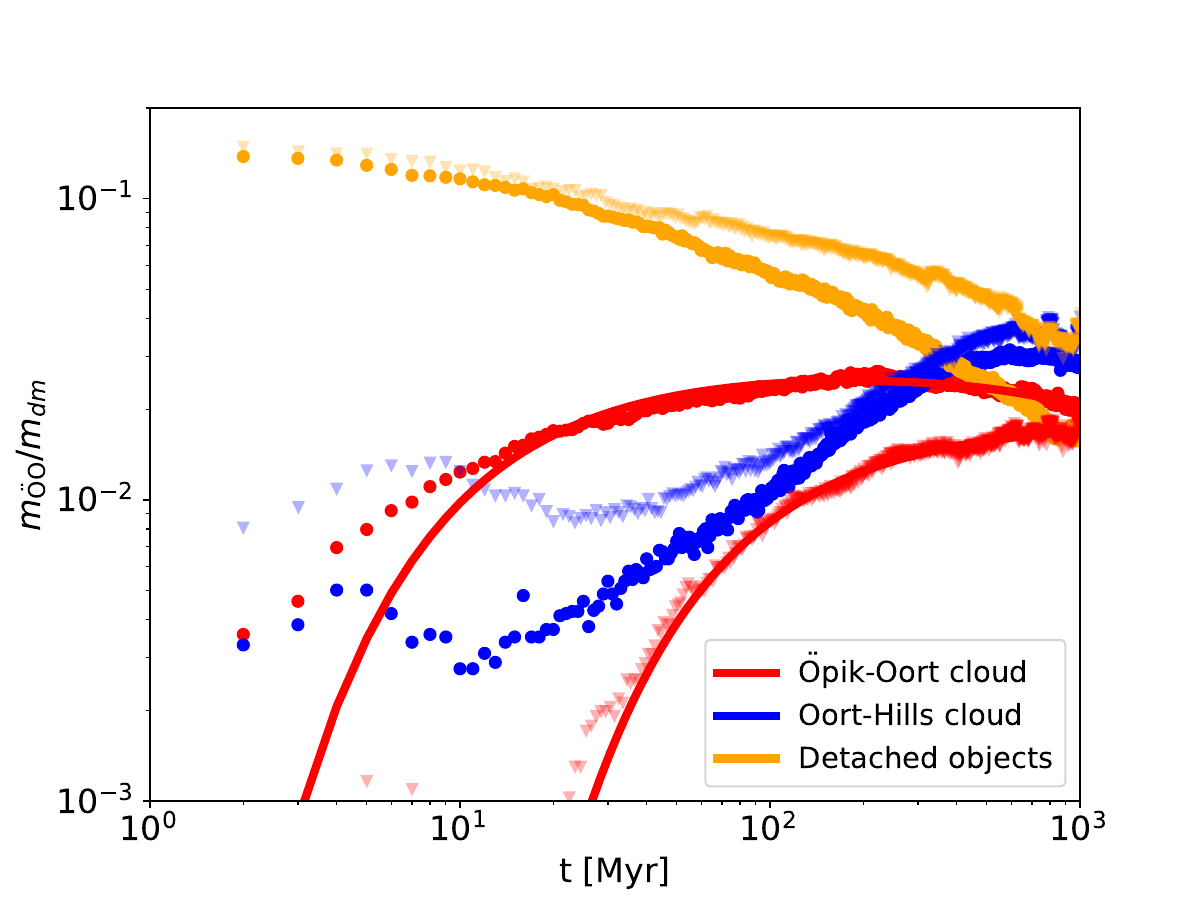}
\caption[]{Relative mass evolution of the Oort cloud and detached
population. For isolated clusters the mass evolution of the \OOc,
\OHc, and the detached populations are given by the upper red, blue
and orange symbols, respectively. The triangles give the mass
evolution in the simulation where the Solar system left the cluster
20\,Myr after the formation of the giant planets, the bullets for
the isolated Solar system. Here we adopted the inner \OOc\, boundary
of $r_{\rm inner} = 30\,000$\,au. The red symbols are identical to
the blue and red dots in the main paper's
\cref{fig:OCGalacticOrbit}. The two solid red curves represent the
fits to the two \OOc\,populations.
\label{fig:OC_evolution}
}
\end{figure}

Ideally, we would like to infer the initial Oort-cloud mass by
calculating backward in time in order to reconstruct when the Sun
left the parent cluster. Since this inversion problem is not
possible, we fit the measured Oort-cloud mass evolution and invert the
fitted curve. We fit a fast-rise-exponential-decay (FRED) function of
the form
\begin{equation}
{\rm FRED} = \mu_{\textrm{\"OO}} \sqrt{e^{t_{\rm rise}/t_{\rm decay}}}
e^{-t_{\rm rise}/t - t/t_{\rm decay}},
\label{Eq:FRED}
\end{equation}
to the mass evolution of the Oort cloud. The three free parameters in
this function includes the proportionality constant
$\mu_{\textrm{\"OO}}$, and the two timescales $t_{\rm rise}$ and
$t_{\rm decay}$. The mass of the Oort cloud then peaks at
$\sqrt{t_{\rm rise} t_{\rm decay}}$. We fitted on the accumulated data
for all 27 runs (see \cref{fig:OC_evolution}). The \OOc\, is best
described with a FRED with a relative peak value of
$\mu_{\textrm{\"OO}} = 0.027$, $t_{\rm rise} = 10$\,Myr, and $t_{\rm
  decay}$ between 4\,000\,Myr and 13\,000\,Myr, whereas for \HOc\,
$\mu_{\textrm{OH}} = 0.027$, $t_{\rm rise} = 80$\,Myr for a similar
decay timescale. The resulting fits for the \OOc\, are presented in
the main paper \cref{fig:OCGalacticOrbit}, \cref{fig:OCmassevolution},
and \cref{fig:OC_evolution}. In \cref{fig:OC_evolution} we present the
mass evolution of various Oort cloud populations as a function of
time. The \OOc\, for both simulations are over-plotted with the fit.

Our adopted choice of the inner boundary for the \OOc\, of $r_{\rm
inner} = 30\,000$\,au is somewhat arbitrary. We therefore fit
$\mu_{\textrm{\"OO}}$ and $t_{\rm rise}$ as a function of this inner
boundary. The values for $m_{\textrm{\"OO}}$ and $t_{\rm rise}$
depend on the inner edge $r_{\rm inner}$ as follows
\begin{equation}
\mu_{\textrm{\"OO}} \simeq 0.066 -0.013 \left( r_{\rm inner}/10\,000{\rm au} \right)
\end{equation}
The rise-time depends on the \OOc\, inner edge as follows
\begin{equation}
t_{\rm rise} \simeq 38\,{\rm Myr} - 9.5\,{\rm Myr} \left( r_{\rm inner}/10\,000\,{\rm au}\right).
\end{equation}

The \OOc\, in the isolated Solar system, starts growing from the start
of the simulation. We anticipate that our starting point coincides
with the moment the giant planets stop migrating and have reached
their current mass. In the clustered environment, the \OOc\, can only
start to form after the Sun escapes and becomes isolated, (in the
simulation this happens at an age of 20\,Myr). This moment is visible
in the dip in the formation of the \OHc, of \cref{fig:OC_evolution}
around 20\,Myr (see blue triangles). In contrast to the \OOc\, The
\OHc\, already starts to be populated while the Solar system is a
cluster member. The exchange of planetesimals between the \OHc\,
(blue) and the detached population (yellow) proceeds differently if
the Solar system is born in isolation (bullets in
\cref{fig:OC_evolution}), or in a clustered environment (triangles).

\begin{figure}
\center
\includegraphics[width=\columnwidth]{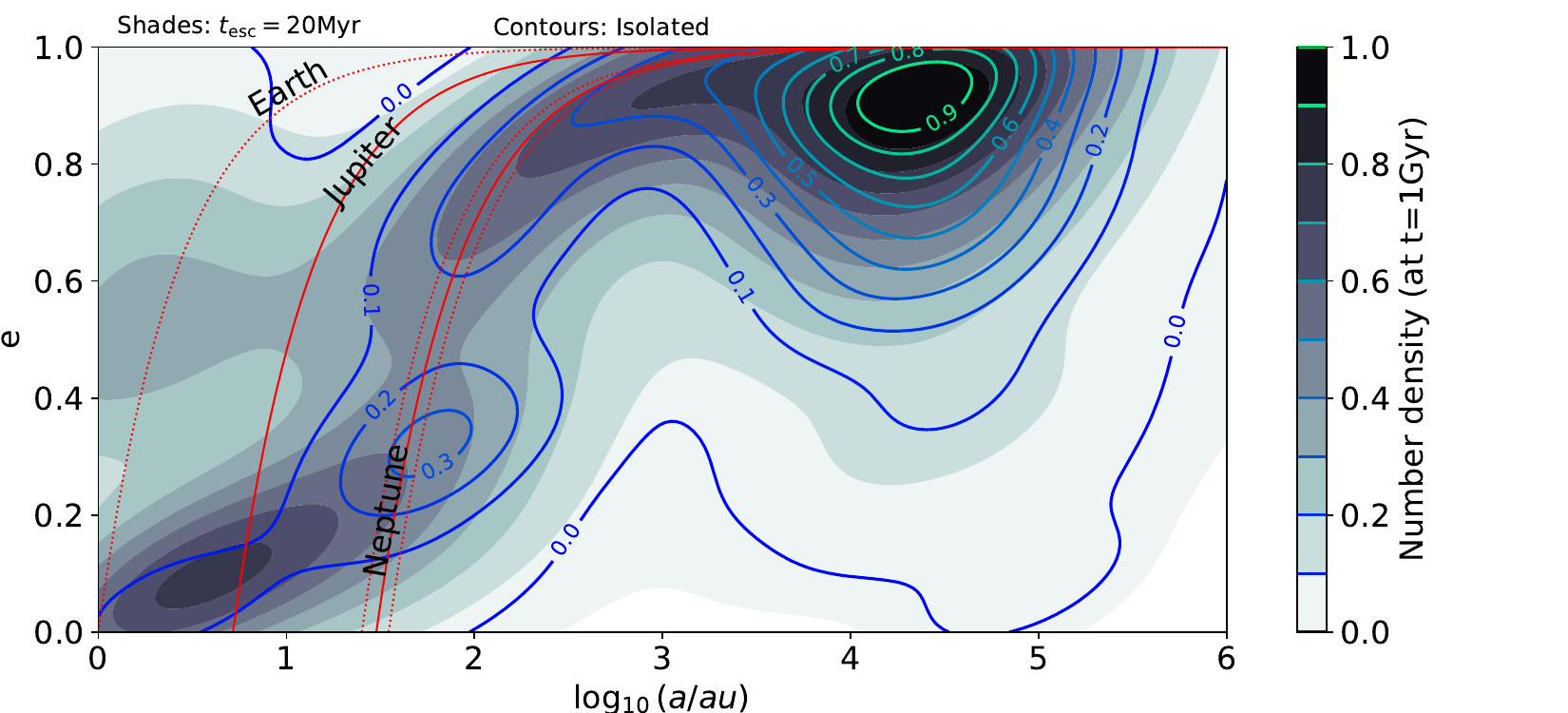}
\caption[]{Semi-major axis v.s. eccentricity of the Solar system
planetesimals at the age of 1000\,Myr for the isolated case
(contours), and if the Solar system left the cluster at an age of
20\,Myr (shades). Both distributions are normalized independently to
the maximum value with 10 equally-spaced contour levels each. In
\cref{fig:S2_ae_at_226Myr} (main paper), we present the results of
the same simulations but at an age of 226\,Myr.
\label{fig:Iso_and_S2Iso_ae_at_1000Myr}
}
\end{figure}

Finally, in \cref{fig:Iso_and_S2Iso_ae_at_1000Myr} we present the
orbital distribution (semi-major axis and eccentricity) of the
planetesimals that remained in the Solar system at an age of
1\,Gyr. The shaded regions indicate the density distribution if the
Solar system escapes the cluster at the age of 20\,Myr, and the
contours for the isolated Solar system.

\end{appendix}

\end{document}